\documentclass[
reprint,
superscriptaddress,
showpacs,preprintnumbers,
 amsmath,amssymb,
 aps,nofootinbib,
]{revtex4-1}

\usepackage{graphicx} 
\usepackage{dcolumn}
\usepackage{bm}
\newcommand{\etap}{\eta^\prime}
\newcommand{\beamE}{2499.1 \pm 2.0}
\newcommand{\beamEc}{1621.6 \pm 0.8}
\newcommand{\targetT}{4.115 \pm 0.001}
\newcommand{\targetCDD}{1.027 \pm 0.002}

\newcommand{\intensityB}{10^{10}}
\newcommand{\reaction}{(p,d)}

\newcommand{\xdelta}{3.51}
\newcommand{\Erange}{\sim 35}

\newcommand{\resolvP}{3.8 \times 10^3}

\newcommand{\C}[1]{{}^{#1}{\rm C}}
\begin{document}
\title{\boldmath Measurement of excitation spectra in the ${}^{12}{\rm C}\reaction$ reaction\\ near the $\etap$ emission threshold}
\author{Y.~K.~Tanaka}
\altaffiliation[Present address: ]{GSI Helmholtzzentrum f\"ur Schwerionenforschung GmbH, Planckstra\ss e 1, 64291 Darmstadt, Germany}
\affiliation{The University of Tokyo, 7-3-1 Hongo, Bunkyo, 113-0033 Tokyo, Japan}
\author{K.~Itahashi}
\email{E-mail: itahashi@riken.jp}
\affiliation{Nishina Center for Accelerator-Based Science, RIKEN, 2-1 Hirosawa, Wako, 351-0198 Saitama, Japan}
\author{H.~Fujioka}
\email{E-mail: fujioka@scphys.kyoto-u.ac.jp}
\affiliation{Kyoto University, Kitashirakawa-Oiwakecho, Sakyo-ku, 606-8502 Kyoto, Japan}
\author{Y.~Ayyad}
\affiliation{RCNP, Osaka University, 10-1 Mihogaoka, Ibaraki, 567-0047 Osaka, Japan}
\author{J.~Benlliure}
\affiliation{Universidade de Santiago de Compostela, 15782 Santiago de Compostela, Spain}
\author{K.-T.~Brinkmann}
\affiliation{Universit\"{a}t Giessen, Heinrich-Buff-Ring 16, 35392 Giessen, Germany}
\author{S.~Friedrich}
\affiliation{Universit\"{a}t Giessen, Heinrich-Buff-Ring 16, 35392 Giessen, Germany}
\author{H.~Geissel}
\affiliation{Universit\"{a}t Giessen, Heinrich-Buff-Ring 16, 35392 Giessen, Germany}
\affiliation{GSI Helmholtzzentrum f\"ur Schwerionenforschung GmbH, Planckstra\ss e 1, 64291 Darmstadt, Germany}
\author{J.~Gellanki}
\affiliation{KVI-CART, University of Groningen, Zernikelaan 25, 9747 AA Groningen, the Netherlands}
\author{C.~Guo}
\affiliation{Beihang University, Xueyuan Road 37, Haidian District, 100191 Beijing, China}
\author{E.~Gutz}
\affiliation{Universit\"{a}t Giessen, Heinrich-Buff-Ring 16, 35392 Giessen, Germany}
\author{E.~Haettner}
\affiliation{GSI Helmholtzzentrum f\"ur Schwerionenforschung GmbH, Planckstra\ss e 1, 64291 Darmstadt, Germany}
\author{M.~N.~Harakeh}
\affiliation{KVI-CART, University of Groningen, Zernikelaan 25, 9747 AA Groningen, the Netherlands}
\author{R.~S.~Hayano}
\affiliation{The University of Tokyo, 7-3-1 Hongo, Bunkyo, 113-0033 Tokyo, Japan}
\author{Y.~Higashi}
\affiliation{Nara Women's University, Kita-Uoya Nishi-Machi, 630-8506 Nara, Japan}
\author{S.~Hirenzaki}
\affiliation{Nara Women's University, Kita-Uoya Nishi-Machi, 630-8506 Nara, Japan}
\author{C.~Hornung}
\affiliation{Universit\"{a}t Giessen, Heinrich-Buff-Ring 16, 35392 Giessen, Germany}
\author{Y.~Igarashi}
\affiliation{KEK, 1-1 Oho, Tsukuba, 305-0801 Ibaraki, Japan}
\author{N.~Ikeno}
\affiliation{Tottori University, 4-101 Koyamacho-minami, 680-8551 Tottori, Japan}
\author{M.~Iwasaki}
\affiliation{Nishina Center for Accelerator-Based Science, RIKEN, 2-1 Hirosawa, Wako, 351-0198 Saitama, Japan}
\author{D.~Jido}
\affiliation{Tokyo Metropolitan University, 1-1 Minami-Osawa, Hachioji, 192-0397 Tokyo, Japan}
\author{N.~Kalantar-Nayestanaki}
\affiliation{KVI-CART, University of Groningen, Zernikelaan 25, 9747 AA Groningen, the Netherlands}
\author{R.~Kanungo}
\affiliation{Saint Mary's University, 923 Robie Street, Halifax, Nova Scotia B3H 3C3, Canada}
\author{R.~Kn\"obel}
\affiliation{Universit\"{a}t Giessen, Heinrich-Buff-Ring 16, 35392 Giessen, Germany}
\affiliation{GSI Helmholtzzentrum f\"ur Schwerionenforschung GmbH, Planckstra\ss e 1, 64291 Darmstadt, Germany}
\author{N.~Kurz}
\affiliation{GSI Helmholtzzentrum f\"ur Schwerionenforschung GmbH, Planckstra\ss e 1, 64291 Darmstadt, Germany}
\author{V.~Metag}
\affiliation{Universit\"{a}t Giessen, Heinrich-Buff-Ring 16, 35392 Giessen, Germany}
\author{I.~Mukha}
\affiliation{GSI Helmholtzzentrum f\"ur Schwerionenforschung GmbH, Planckstra\ss e 1, 64291 Darmstadt, Germany}
\author{T.~Nagae}
\affiliation{Kyoto University, Kitashirakawa-Oiwakecho, Sakyo-ku, 606-8502 Kyoto, Japan}
\author{H.~Nagahiro}
\affiliation{Nara Women's University, Kita-Uoya Nishi-Machi, 630-8506 Nara, Japan}
\author{M.~Nanova}
\affiliation{Universit\"{a}t Giessen, Heinrich-Buff-Ring 16, 35392 Giessen, Germany}
\author{T.~Nishi}
\affiliation{Nishina Center for Accelerator-Based Science, RIKEN, 2-1 Hirosawa, Wako, 351-0198 Saitama, Japan}
\author{H.~J.~Ong}
\affiliation{RCNP, Osaka University, 10-1 Mihogaoka, Ibaraki, 567-0047 Osaka, Japan}
\author{S.~Pietri}
\affiliation{GSI Helmholtzzentrum f\"ur Schwerionenforschung GmbH, Planckstra\ss e 1, 64291 Darmstadt, Germany}
\author{A.~Prochazka}
\affiliation{GSI Helmholtzzentrum f\"ur Schwerionenforschung GmbH, Planckstra\ss e 1, 64291 Darmstadt, Germany}
\author{C.~Rappold}
\affiliation{GSI Helmholtzzentrum f\"ur Schwerionenforschung GmbH, Planckstra\ss e 1, 64291 Darmstadt, Germany}
\author{M.~P.~Reiter}
\affiliation{GSI Helmholtzzentrum f\"ur Schwerionenforschung GmbH, Planckstra\ss e 1, 64291 Darmstadt, Germany}
\author{J.~L.~Rodr\'{i}guez-S\'{a}nchez}
\affiliation{Universidade de Santiago de Compostela, 15782 Santiago de Compostela, Spain}
\author{C.~Scheidenberger}
\affiliation{Universit\"{a}t Giessen, Heinrich-Buff-Ring 16, 35392 Giessen, Germany}
\affiliation{GSI Helmholtzzentrum f\"ur Schwerionenforschung GmbH, Planckstra\ss e 1, 64291 Darmstadt, Germany}
\author{H.~Simon}
\affiliation{GSI Helmholtzzentrum f\"ur Schwerionenforschung GmbH, Planckstra\ss e 1, 64291 Darmstadt, Germany}
\author{B.~Sitar} 
\affiliation{Comenius University Bratislava, Mlynsk\'{a} dolina, 842 48 Bratislava, Slovakia}
\author{P.~Strmen}
\affiliation{Comenius University Bratislava, Mlynsk\'{a} dolina, 842 48 Bratislava, Slovakia}
\author{B.~Sun}
\affiliation{Beihang University, Xueyuan Road 37, Haidian District, 100191 Beijing, China}
\author{K.~Suzuki}
\affiliation{Stefan-Meyer-Institut f\"{u}r subatomare Physik, Boltzmangasse 3, 1090 Vienna, Austria}
\author{I.~Szarka}
\affiliation{Comenius University Bratislava, Mlynsk\'{a} dolina, 842 48 Bratislava, Slovakia}
\author{M.~Takechi}
\affiliation{Niigata University, 8050 Ikarashi 2-no-cho, Nishi-ku, 950-2181 Niigata, Japan}
\author{I.~Tanihata} 
\affiliation{RCNP, Osaka University, 10-1 Mihogaoka, Ibaraki, 567-0047 Osaka, Japan}
\affiliation{Beihang University, Xueyuan Road 37, Haidian District, 100191 Beijing, China}
\author{S.~Terashima}
\affiliation{Beihang University, Xueyuan Road 37, Haidian District, 100191 Beijing, China}
\author{Y.~N.~Watanabe}
\affiliation{The University of Tokyo, 7-3-1 Hongo, Bunkyo, 113-0033 Tokyo, Japan}
\author{H.~Weick}
\affiliation{GSI Helmholtzzentrum f\"ur Schwerionenforschung GmbH, Planckstra\ss e 1, 64291 Darmstadt, Germany}
\author{E.~Widmann}
\affiliation{Stefan-Meyer-Institut f\"{u}r subatomare Physik, Boltzmangasse 3, 1090 Vienna, Austria}
\author{J.~S.~Winfield}
\affiliation{GSI Helmholtzzentrum f\"ur Schwerionenforschung GmbH, Planckstra\ss e 1, 64291 Darmstadt, Germany}
\author{X.~Xu}
\affiliation{Universit\"{a}t Giessen, Heinrich-Buff-Ring 16, 35392 Giessen, Germany}
\affiliation{GSI Helmholtzzentrum f\"ur Schwerionenforschung GmbH, Planckstra\ss e 1, 64291 Darmstadt, Germany}
\author{H.~Yamakami}
\affiliation{Kyoto University, Kitashirakawa-Oiwakecho, Sakyo-ku, 606-8502 Kyoto, Japan}
\author{J.~Zhao}
\affiliation{Beihang University, Xueyuan Road 37, Haidian District, 100191 Beijing, China}

\collaboration{$\eta$-PRiME/Super-FRS Collaboration}

\date{\today}

\begin{abstract}
Excitation spectra of $\C{11}$ were measured in the
$\C{12}\reaction$ reaction near the $\etap$ emission threshold.
A proton beam extracted from the synchrotron SIS-18 at GSI with an incident energy of 2.5 GeV
impinged on a carbon target. The momenta of deuterons emitted at $0^\circ$
were precisely measured with the fragment separator FRS operated as
a spectrometer. In contrast to theoretical predictions on the
possible existence of deeply bound $\etap$ mesic states in carbon nuclei, 
no distinct structures 
were observed associated with the formation of bound states.
The spectra were analyzed to set stringent constraints
on the formation cross section and on the hitherto barely-known 
$\etap$-nucleus interaction.
\end{abstract}


\maketitle
The mass of the $\etap$ meson is exceptionally large (958 MeV/$c^2$)
compared with other mesons in the same pseudo-scalar multiplet. This
large mass is known as the ``U(1) problem'' raised by 
Weinberg~\cite{Weinberg1975}. 
Since the origin of the exceptionally large mass is attributed to
the effect of the $U_A(1)$ anomaly in QCD (Quantum ChromoDynamics) under
the presence of spontaneous breaking of chiral symmetry, it is natural
to expect a weakening of such an anomaly effect in a nuclear medium,
where chiral symmetry may partially be restored~\cite{Hayano2010,Leupold2010}.
This suppression of the anomaly effect would lead to a large mass
reduction in a nuclear medium~\cite{Jido2012}. Recent model calculations have
predicted a 4--15\% mass reduction 
at normal nuclear density~\cite{Costa2005,Nagahiro2006, Bass2006, Sakai2013}. Thus, 
the observation of such a modification would provide novel insights into
strongly-interacting many-body systems and the QCD vacuum.

A mesic atom or nucleus, which is a bound state of a meson and a nucleus,
will offer a testing ground for the investigation of in-medium meson properties
which can be different from those in the QCD vacuum
due to partial restoration of chiral symmetry~\cite{Hayano2010, Leupold2010}.
For example, deeply-bound pionic atoms in nuclei, in which a modification of the 
isovector part of the s-wave pion-nucleus potential manifests itself,
are very well-established systems~\cite{Yamazaki2012,Suzuki:2002ae}.
Distinct peak structures corresponding to pionic states
with different configurations, especially pions in the atomic $1s$ state,
can be identified in the missing-mass spectrum of the ($d$,$^3\mathrm{He}$) reaction~\cite{Gilg1999ua}.
In contrast to invariant-mass spectroscopy of
hadrons decaying into multi-particle final states
in a nuclear medium, missing-mass spectroscopy is an alternative approach 
providing experimental access to in-medium meson properties.

In addition a quantum many-body system, such as an atomic nucleus, exhibits various phenomena
which cannot be observed in a two-body system. The strong interaction is indeed ``strong''
in that a typical binding energy of a particle relative to its rest energy is large,
in contrast to the electro-magnetic interaction which binds an electron and an atomic nucleus.
Hence, the sub-threshold behavior of two-body interactions is of importance in building up a 
many-body system. This may be also the case for the $\etap$-nucleon interaction. 

The mass reduction of the $\etap$ meson leads to an attractive interaction
between an $\eta'$ meson and a nucleus. 
Hence, an experimental observation of $\eta'$-nucleus bound states
may provide a direct clue to deduce the $\eta'$ mass at normal nuclear density.
It should be noted that a clear distinction of neighboring levels in 
a spectrum requires a rather narrow absorption width.
Numerical results of an $\eta'$-nucleus optical potential based on a chiral unitary model
show that the depth of the imaginary part of the potential (half absorption width) is much smaller 
than that of the real part (mass reduction), regardless of the strength of the two-body $\eta'N$ interaction~\cite{Nagahiro2012}.

Experimental information on low-energy $\eta'$-nucleus interaction is very limited. 
A recent measurement of $\eta'$ photoproduction and the emission from nuclei at CBELSA/TAPS
was utilized to evaluate the $\eta'$-nucleus optical potential for the 
first time~\cite{Nanova2013}. The absorption width, 
for the average $\eta'$ momentum of $1050\,\mathrm{MeV}/c$, was deduced 
to be 15--25$\,\mathrm{MeV}$ at normal nuclear density, from a 
measurement of the transparency ratio of $\etap$ mesons propagating 
in various nuclear targets~\cite{Nanova2012}.
Furthermore, the excitation function and the momentum distribution of $\eta'$ mesons, produced and emitted from $^{12}\mathrm{C}$, 
were compared with a model calculation with the real part of the potential as a free parameter leading to 
a potential depth of 
$-37\pm 10 \mathrm{(stat)}\pm 10\mathrm{(syst)}\,\mathrm{MeV}$~\cite{Nanova2013}. 
In contrast, a small $\eta'N$ scattering length
with a real part consistent with zero was 
obtained from the excitation function of 
the $pp\to pp\eta'$ reaction very close to the threshold~\cite{Czerwinski2014}.

As can be inferred from the $\eta$-nucleon and
$\eta$-nucleus interactions, which have been investigated in a variety of reactions
~\cite{Machner2015,Haider2015},
it is far from straightforward to extrapolate the on-shell scattering amplitude to the subthreshold region, 
pointing to the importance of meson-nucleus bound systems, serving as a unique probe of subthreshold behavior.

A direct measurement of the in-medium mass reduction was proposed by
spectroscopy of $\etap$ bound states in a nucleus
by measuring an excitation spectrum of $\C{11}$ in the 
${}^{12}{\rm C}\reaction$ reaction near
the $\etap$ emission threshold~\cite{Itahashi2012}.
Theoretical calculations indicated the possible existence of 
such bound states. One based on Nambu--Jona-Lasinio model predicted a
sufficiently attractive potential of $-150$~MeV~\cite{Costa2005,Nagahiro2006}
to accommodate $\etap$ in a carbon nucleus provided that the absorption is not too strong. 
Considerations on the momentum transfer and the
differential cross section of the elementary $n\reaction\etap$ reaction
led to a preferable incident proton energy of 2.5 GeV.

As discussed in Ref.~\cite{Itahashi2012}, excitation energy spectra were predicted 
for different assumed $\etap$-nucleus interactions
expressed by attractive real and absorptive imaginary potential 
depths $(V_0, iW_0)$ MeV at the nuclear center.
The predicted spectra 
demonstrated the experimental sensitivity for the bound states
and the advantage of an unbiased spectral analysis
in an inclusive measurement that surpasses 
the disadvantages of a small signal-to-noise ratio
mainly arising from large cross sections of multiple pion production in the reaction.
Thus, we aimed at a region where attraction is relatively strong and absorption is weak
by achieving extremely good statistics of $\leq 1\%$ relative errors
and a moderate resolution of $\leq 10$\ MeV over the spectral 
region of interest.

The ${}^{11}{\rm C}$ excitation spectra were measured near the
$\etap$ emission threshold using the ${}^{12}{\rm C}\reaction$ reaction
at zero degrees. We employed a proton beam with the 
energy of $\beamE$\ MeV extracted from 
the synchrotron SIS-18 at GSI, Darmstadt, 
impinging on a carbon target of natural isotopic composition 
with the thickness of $\targetT$ g/cm$^2$. The beam
intensity was $\sim \intensityB$/s and the
spill length and the cycle were 4 s and 7 s, respectively.
The intensity was directly measured 
with an uncertainty of $5.5$\% 
by inserting a detector SEETRAM~\cite{Jurado2002} in the beam near the target.
The typical horizontal beam spot size was 1 mm.

\begin{figure}
\includegraphics[width=85mm]{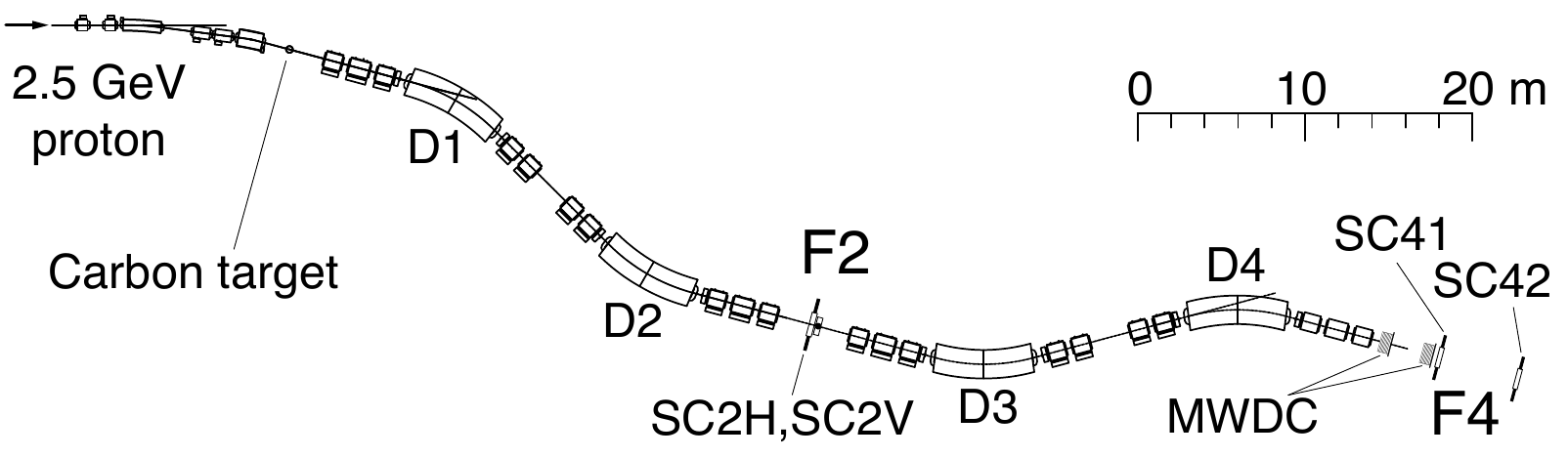}
\vskip-3mm
\caption{\label{Fig: setup}A schematic view of the FRS used as a spectrometer
and detectors used in the present analysis. A 2.5 GeV proton beam impinged on a carbon target.
Deuterons emitted 
in the $\C{12}\reaction$ reaction were momentum-analyzed at F4 
and the tracks were measured
by MWDCs. Sets of 5 mm-thick plastic 
scintillation counters (SC2H, 2V, and SC41) and 20 mm-thick one
(SC42) were installed at F2 and F4 for TOF measurement.}
\end{figure}

The emitted deuterons had a momentum of $2814.4$\ MeV/$c$ at the $\etap$ production
threshold
after energy loss in the target and were momentum-analyzed by the 
FRS~\cite{Geissel:1991zn} used as a spectrometer 
with a specially developed ion optical setting. The central focal plane (F2) 
was momentum-achromatic and the final focal plane (F4) was dispersive
with a designed momentum resolving power of $\resolvP$.
The F4 dispersion was measured to be $\xdelta$\ cm/\%. 
The deuteron tracks were measured by 
two sets of MWDCs (multi-wire drift chambers) separately installed at
a distance of about one meter
near F4 as depicted in Fig.~\ref{Fig: setup}.

Four sets of plastic scintillation counters installed
at F2 and F4 identified the particles by measuring
the time-of-flight (TOF) for signal deuterons ($\sim$ 150 ns) 
and for background protons ($\sim$ 132 ns). Count rates at F4 during the 
spill extraction were $\sim 250$ kHz for protons and $\sim 1$ kHz for deuterons. 
99.5\% of the protons were rejected in the TOF based triggers
while $\sim 100$\% of deuterons were selected.
The DAQ live rate ranged between 30--40\% and the 
recorded trigger rate varied around 1 kHz during the spill extraction.
After further selection based on the waveform analyses of PMT signals of the 
scintillation counters rejecting multiple hits, 
we achieved $\leq 4\%$ deuteron overkill in the analysis with 
a very small $\sim 2 \times 10^{-4}$ proton contamination fraction of 
the deuteron identified events.

During the measurement, we accumulated data 
to cover a wider energy region between $-91$ and $+34$ MeV 
around the $\etap$ emission threshold by scaling all FRS 
magnets with seven factors between 0.980 and 1.020.
An excitation energy range of $\Erange$\ MeV was covered in one setting 
by the central acceptance region, where one could safely rely on 
the momentum acceptance of the spectrometer. The momentum acceptance was determined
by a code MOCADI~\cite{Iwasa:1997bm}
simulating the particle tracks based on the 
geometries of the magnets, the ion optical transport,
the apertures of the beam pipes, effects of the materials, 
and the detector performances. The uncertainties in the acceptance estimation
were taken into account in the subsequent analyses.

The central momenta of the seven settings were calibrated by using an elastic D$\reaction$ reaction on 
a deuterated polyethylene (CD$_2$) target
with the thickness of $\targetCDD$ g/cm$^2$ 
using a $\beamEc$\ MeV proton beam. 
Nearly mono-energetic deuterons with the momentum of $2828.0 \pm 1.0$\ MeV/$c$ 
were emitted forward within the solid angle 
of the FRS (horizontal: $\pm 15$ mrad and vertical: $\pm 20$ mrad).
We also evaluated the ion optical aberrations from the measured 
correlations between positions and angles and adopted the corrections.  
The calibration data were taken every eight hours during the 
total 70.5 hours of production measurement and confirmed
the stability of the
spectrometer system.

The momentum spectra 
for seven central-momentum settings were combined
after the acceptance and optical aberration corrections
by fitting the spectra in the overlap regions of 
neighboring settings. Note here
that this procedure decreased the degrees of freedom of the data,
which turned out to cause minor influences as we discuss below.

\begin{figure}[hbtp]
\centering \includegraphics[width=85.0mm]{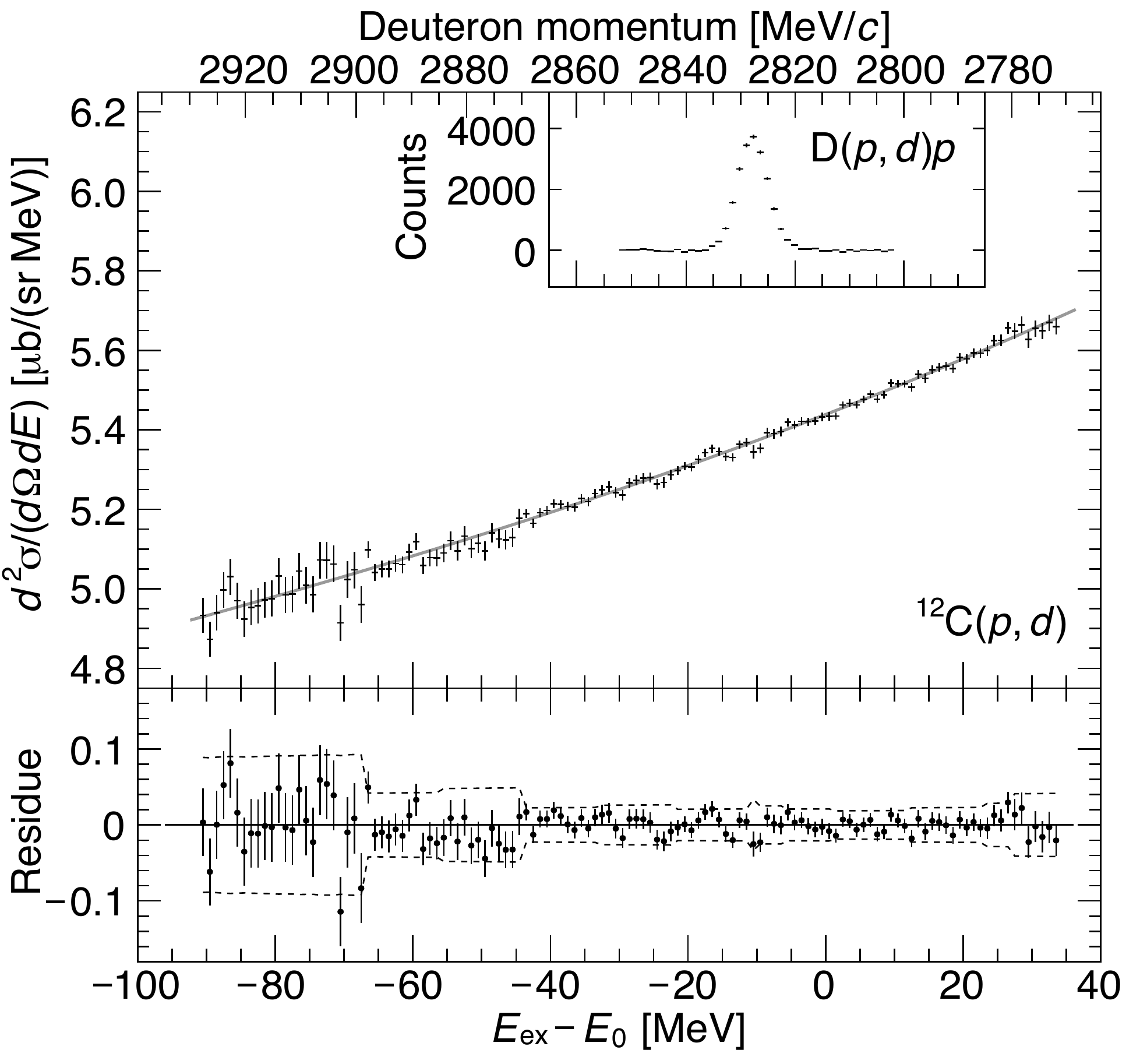}
\caption{\label{fig_spectrum} (top)
Excitation spectrum of $^{11}$C measured in the $^{12}$C($p$,$d$) reaction at a proton energy of 2.5 GeV.
The abscissa is the excitation energy $E_{\mathrm{ex}}$ referring to the $\etap$ emission threshold $E_0 =  957.78$ MeV.
The overlaid gray solid curve displays a fit of the spectrum with a third-order polynomial.
The upper horizontal axis shows the deuteron momentum scale.
The inset presents a deuteron momentum spectrum measured 
in the elastic D$(p,d)p$ reaction using a 1.6 GeV proton beam.
(bottom) Fit residues with envelopes of two standard deviations.
}
\end{figure}

Figure~\ref{fig_spectrum} (top) shows the measured excitation
spectrum of the $\C{12}\reaction$ reaction
near 
the $\eta'$ emission threshold.
The excitation energy $E_{\mathrm{ex}}$ relative 
to the threshold $E_0 = 957.78$~MeV is shown by the bottom axis
and the deuteron momentum by the top axis.
The ordinate is the double differential cross section of the reaction
at zero degrees which has an error of $\pm 13$\% in the absolute scale mainly as a result 
of the uncertainties in the incident beam intensities and in the solid angle.
The systematic error associated with the excitation energy is 
deduced to be 1.7~MeV, mainly owing to the uncertainties of the beam energies.

No distinct narrow structure has been observed 
in the excitation spectrum
in spite of the extremely good statistical sensitivity 
at a level of better than 1\%.
The measured cross section steadily increases from
4.9 to 5.7~$\mu$b/(sr\,MeV) within the measured
excitation energy region from $-91$ to 
$+34$ MeV 
and agrees within an order of magnitude
with the simulated cross sections of the quasi-free processes 
$pN \rightarrow dX (X=2\pi, 3\pi, 4\pi, \omega)$~\cite{Itahashi2012}, where $N$ denotes a nucleon
in a carbon nucleus.
The measured spectrum has been fitted over the whole region
by a third-order polynomial displayed as solid
gray curve where $\chi^2$ and the number of degrees of freedom are 119 
and 121, respectively.
For positive excitation energies, the fit may include increasing contributions from 
quasi-free $\etap$ production, estimated to reach the order of some 10 nb/(sr MeV) 
at $E_{\mathrm{ex}}-E_0 = 30$ MeV~\cite{Nagahiro2013}.
The fitting residues are shown in the bottom panel with envelopes of two 
standard deviations. No significant structure is observed.
Note that the jumps in the envelope reflect
edges between two neighboring spectrometer settings.

The inset displays the measured momentum distribution of 
elastically scattered deuterons
in a calibration measurement D$(p,d)p$ with the CD$_2$ target after subtracting 
the carbon contribution, demonstrating 
the symmetrical response of the momentum measurement.
The energy resolution during the production runs 
has been estimated by fitting the spectrum with a Gaussian
to yield $\sigma_{\mathrm{exp}}=2.5\pm0.1$~MeV over the whole 
energy region after
considering the energy losses in the 
targets and the momentum spreads of the incident beams.

Since no distinct structure has been observed in the spectrum,
we have deduced upper limits of the 
formation cross section of $\eta'$ mesic nuclei 
as a function of the excitation energy $E_{\mathrm{ex}}$ and 
the width $\Gamma$ (FWHM).
We have assumed that the spectrum has two components, a smooth continuous part
that is expressed by a third-order polynomial and a formation cross section of $\etap$-mesic nuclei
given by a Voigt function, i.e. a Lorentzian with the width $\Gamma$ folded by a 
Gaussian with the width of $\sigma_{\mathrm{exp}}$.
The spectrum has been fitted 
by the sum of the two components within a region of $\pm$ 35~MeV around the Lorentzian center
with the height of the Voigt function and four coefficients of the polynomial 
as free parameters.
We have evaluated upper limits at the 95\% confidence level 
assuming the probability density functions to be Gaussian,
integrating them in the physical non-negative regions, and finding points 
where 95\% of the probability-density-function areas have been covered by the regions beneath.

\begin{figure}[hbtp]
\centering \includegraphics[width=85.0mm]{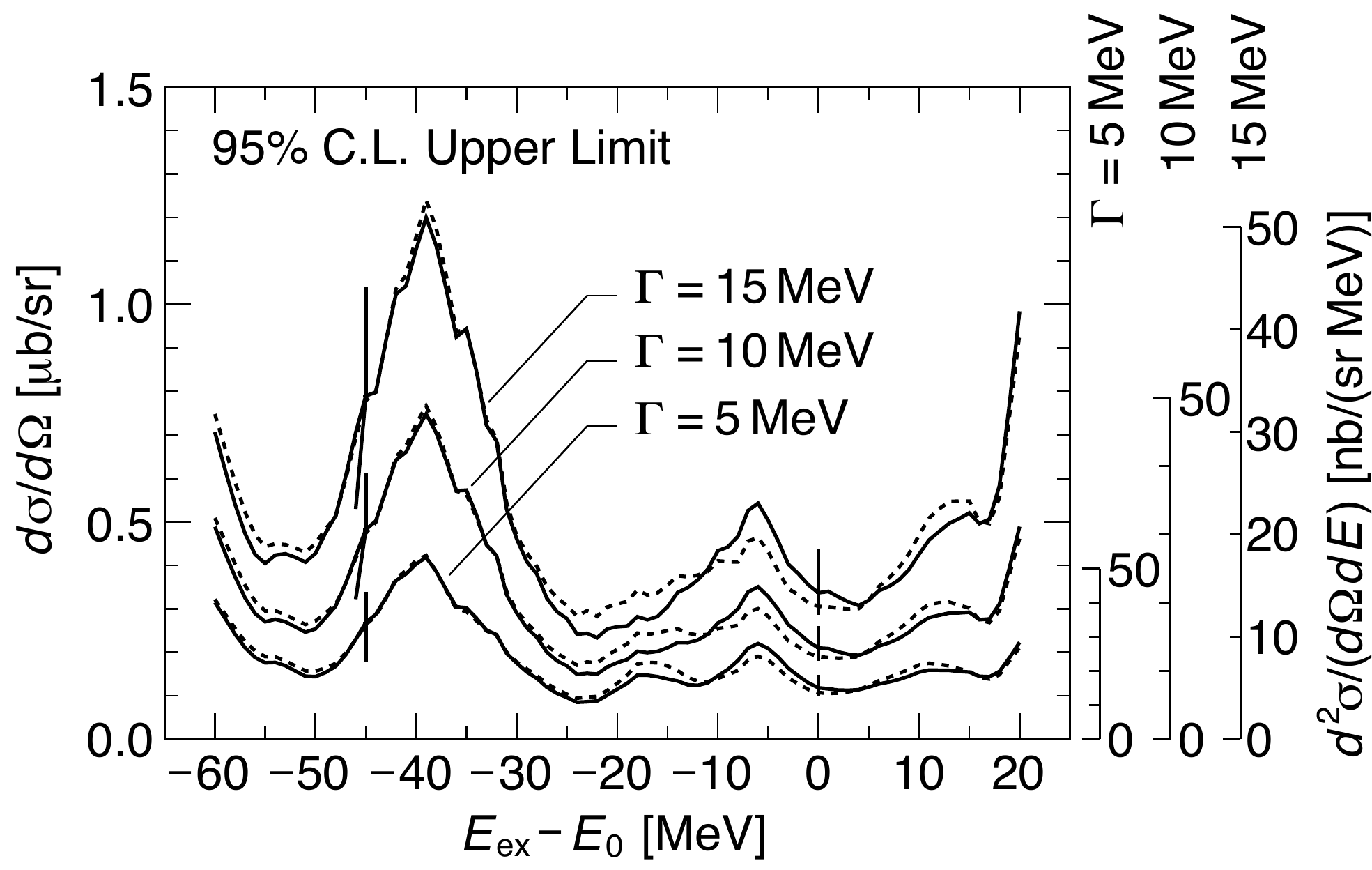}
\caption{\label{fig_upper_limit} 
Upper limits for the formation cross section 
of $\eta'$ mesic nuclei 
at the 95\% confidence levels
evaluated by two methods, namely by fitting the combined spectrum (Fig.~\ref{fig_spectrum})
shown by the solid lines and by simultaneously fitting the spectra for seven 
central-momentum settings (dashed lines).
The limits are presented as functions of the excitation energy $E_{\mathrm{ex}}$ 
for three tested natural widths of $\Gamma=5, 10, 15$~MeV.
The scales to the right indicate the upper limits of the differential cross sections
at the heights of the Lorentzian for each tested natural width. Typical systematic errors
are shown by the vertical bars.
}
\end{figure}

Repeating the above procedure by changing the Lorentzian positions and widths, 
we have obtained upper limits for the region 
$-60 \mathrm{~MeV} \leq E_{\mathrm{ex}}-E_0 \leq +20 \mathrm{~MeV}$ and $\Gamma = 5, 10, 15$~MeV 
as presented by the solid curves in Fig.~\ref{fig_upper_limit}.
The limits in the differential cross section $d\sigma$/$d\Omega$
are indicated by the vertical axis on the left and 
those in the Lorentzian heights $d^2\sigma$/($d\Omega$ $dE$) 
by the three scales on the right side for each $\Gamma$. 
Typical systematic errors are indicated by the vertical bars,
which arise from the uncertainties in the beam intensity, the beam energy, the spectrometer acceptance,
and the spectral resolution, and by moving the fitting boundaries by $\pm$ 5 MeV.
We have set upper limits particularly rigid near the $\eta'$ emission threshold: 
0.1--0.2 $\mu$b/sr for $\Gamma = 5$ MeV,
0.2--0.4 $\mu$b/sr for $\Gamma = 10$ MeV, and
0.3--0.6 $\mu$b/sr for $\Gamma = 15$ MeV.
The above analysis has been checked by a nearly identical procedure on
the separate spectra for the seven spectrometer settings.
Simultaneous fitting on each spectrum has yielded results shown by the
dashed curves in Fig.~\ref{fig_upper_limit},
which are consistent with those in the analysis of the combined spectrum.

The resulting upper limits near the threshold of the peak height 
$\sim 20$ nb/(sr MeV) exclude the existence of 
narrow peak structures with the height as large as 40~nb/(sr MeV)
expected for potential parameters $(V_0, W_0)=(-150,-10)$~MeV~\cite{Nagahiro2013},
which was predicted by a theoretical calculation based on the NJL model~\cite{Nagahiro2006} 
and the absorption width suggested by the measured transparency ratio~\cite{Nanova2012}.

\begin{figure}[hbtp]
\centering \includegraphics[width=85.0mm]{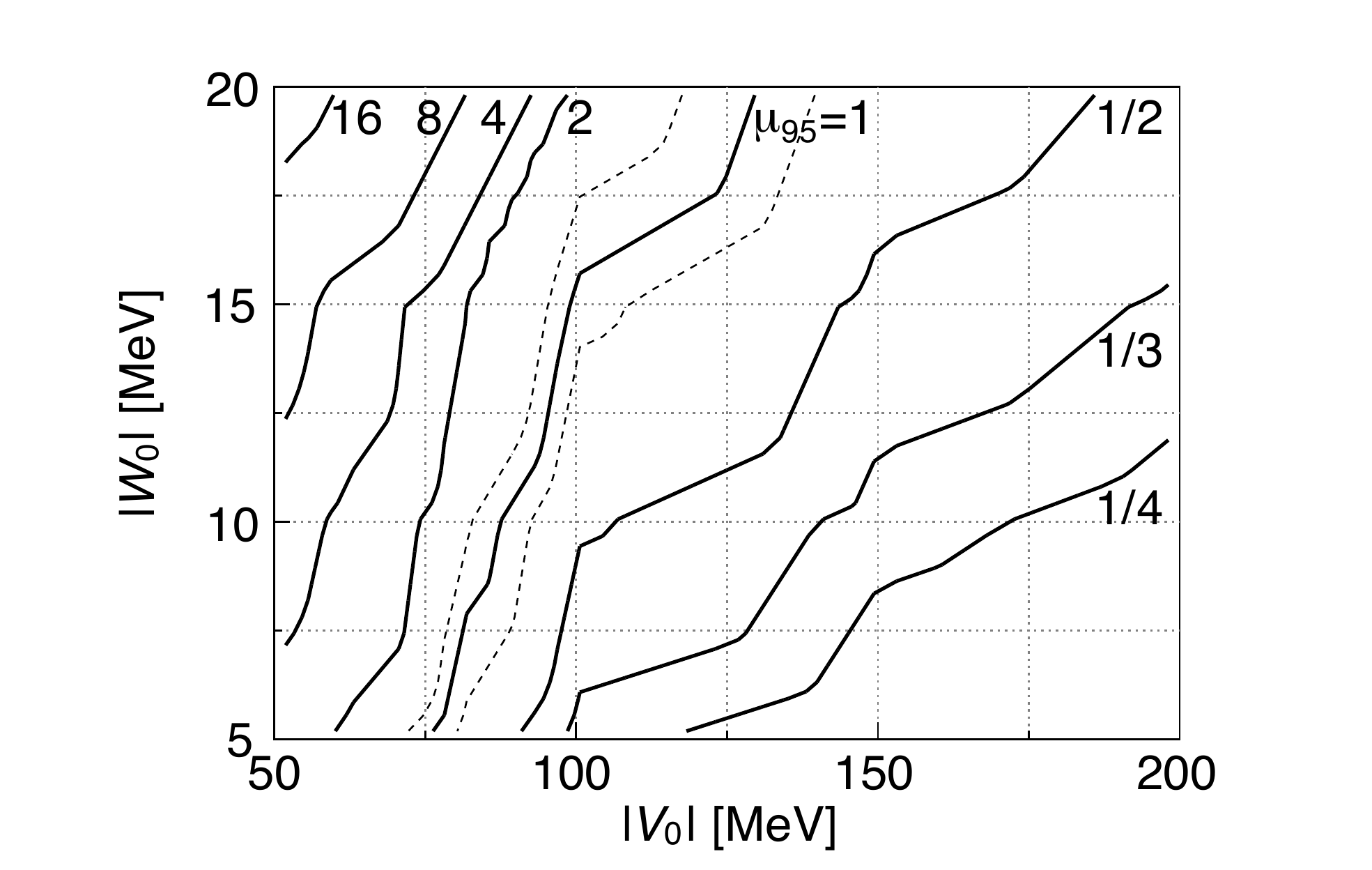}
\caption{\label{fig_potential}
A contour plot of $\mu_{95}$ (solid curves),
upper limit of the scale-parameter $\mu$ at the 95\% C.L.,
on a plane of real and imaginary potential parameters $(|V_0|,|W_0|)$.
The limits have been evaluated for the potential parameter 
combinations $(V_0,W_0)$ in $\{-50,-100,-150,-200\}\times \{-5,-10,-15,-20\}$ and 
$\{-60,-80\}\times \{-5,-10,-15\}$~MeV
and linearly interpolated in-between. Dashed curves show a band of $\mu_{95}=1$ contour
indicating the systematic errors. Regions for $\mu_{95} \leq 1$ are excluded by the present analysis.}
\end{figure}

For further discussion of the constraints set on the $\etap$-nucleus interaction,
we have compared the combined spectrum with theoretical
spectra described in Ref.~\cite{Nagahiro2013} in a space of potential parameters $(|V_0|,|W_0|)$.
For each potential-parameter combination of
$|V_0|=\{50, 100, 150, 200\}$~MeV $\times$ $|W_0|=\{5, 10, 15, 20\}$~MeV and 
$|V_0|=\{60, 80\}$~MeV $\times$ $|W_0|= \{5, 10, 15\}$~MeV,
the spectrum has been fitted in an energy region between $-40$ and $30$ MeV
by a sum of a third-order polynomial and
a theoretical spectrum scaled by a scale-parameter $\mu$ and folded by a Gaussian with the spectral
resolution $\sigma_{\mathrm{exp}}$. Wider ($[-45,35]$ MeV) and narrower ($[-35,25]$ MeV)
fit-regions have also been tested. In a similar way as above, 
upper limits of the scale-parameter $\mu_{95}$ have been evaluated at the 95\% C.L. and
are displayed on a potential-parameter plane in Fig.~\ref{fig_potential} after
linear-interpolation between the calculated points.
The dashed curves show $\mu_{95} = 1$ in a band accounting for the estimated
systematic errors.

Looking closely at Fig.~\ref{fig_potential}, one finds that 
more stringent constraints (expressed by smaller $\mu_{95}$) 
are set for larger $|V_0|$ and smaller $|W_0|$.
Potential-parameter sets giving $\mu_{95}$ smaller than 1
are excluded by the 95\% C.L. within the present analysis.
Note here that theoretical calculations are subject to 
an uncertainty\footnote{From Fig.~3 in Ref.~\cite{Grishina2000}, 
one can read the boundaries in the total cross section of 
$pn \rightarrow d\eta'$ for the c.m.~energy region of interest.
The resultant range is in agreement with a calculation
by K.~Nakayama~\cite{Nakayama2012}.}
of a factor of 2 originating in the estimate of the cross section
of the elementary process $pn \rightarrow d\eta'$ of 30~$\mu$b/sr~\cite{Itahashi2012}, 
which has not yet been determined experimentally,
and the experimental determination is of particular importance.
Thus, the $\mu_{95} = 1/2$ contour, for instance, corresponds to a C.L. 95\% 
upper limit for the case that the absolute theoretical cross section was 
over-estimated by a factor of $1/\mu_{95} = 2$.

In conclusion,
we have conducted a high accuracy measurement of the 
excitation spectrum of the $\C{12}\reaction$ reaction near 
the $\etap$ emission threshold. We accomplished targeted
statistical significance, spectral resolution, and overall
accuracy in the measurement. 
No distinct structures are observed in the spectrum.
Thus, a strongly attractive potential of $V_0 \sim -150$ MeV 
predicted by the NJL model~\cite{Nagahiro2006} is rejected
for a relatively shallow imaginary potential of $|W_0| \sim 10$ MeV~\cite{Nanova2012}.
The present experiment has only limited sensitivity 
for relatively weak attraction implied by the $\etap N$ scattering length~\cite{Czerwinski2014}
and suggested by the $\etap$ photoproduction experiments~\cite{Nanova2013}.
In the near future, we will extend the experimental sensitivity by 
constructing a detector system to efficiently
select events originating from the formation of $\etap$-mesic nuclei
by tagging the decay particles in an experiment at GSI/FAIR.
We also consider possibilities of using other reaction channels
such as $(\pi,N)$.

The authors acknowledge 
the GSI staffs for their support and staffs of
Institut f\"ur Kernphysik, Forschungszentrum J\"ulich
for a test at COSY. We also thank Dr.~M.~Tabata for a specially developed aerogel.
Y.K.T. acknowledges financial support
from Grant-in-Aid for JSPS Fellows (No.\ 258155) and 
H.F. from Kyoto University Young Scholars Overseas Visit Program.
This work is partly supported by a MEXT Grants-in-Aid for Scientific 
Research on Innovative Areas (Nos.\ JP24105705 and JP24105712),
JSPS Grants-in-Aid for Scientific Research (S) (No.\ JP23224008)
and for Young Scientists (A) (No.\ JP25707018),
by the National Natural Science Foundation of China (No.~11235002)
and by the Bundesministerium f\"ur Bildung und Forschung.

\end{document}